\documentclass{article}
\usepackage{spconf,amsmath,graphicx}
\usepackage{cite}
\usepackage{amssymb,amsfonts}
\usepackage{algorithmic}
\usepackage{textcomp}
\usepackage{xcolor}
\usepackage{url}
\usepackage{pgfplots}
\usepackage{caption}
\usepackage[labelformat=simple]{subcaption}

\usepackage{multirow}
\usepackage{hyperref}
\usepgfplotslibrary{groupplots,dateplot}
\renewcommand{\thetable}{\Roman{table}}
\usepackage{fancyhdr}

\fancypagestyle{FirstPage}{
	\lfoot{
		\fbox{
			\parbox{\textwidth}{%
				\footnotesize \textcopyright 2024 IEEE. Personal use of this material is permitted. Permission from IEEE must be obtained for all other uses, in any current or future media, including reprinting/republishing this material for advertising or promotional purposes, creating new collective works, for resale or redistribution to servers or lists, or reuse of any copyrighted component of this work in other works.}
	}}
}

\title{A Study on the Effect of Color Spaces in Learned Image Compression}
\name{Srivatsa Prativadibhayankaram$^{\dagger, \mathsection}$, Mahadev Prasad Panda$^{\dagger}$, Jürgen Seiler$^{\mathsection}$, Thomas Richter$^{\dagger}$}
\secondlinename{Heiko Sparenberg$^{\dagger,\P}$, Siegfried Foessel$^{\dagger}$, Andr\'{e} Kaup$^{\mathsection}$}
\address{$^{\dagger}$ \textit{Moving Picture Technologies}, Fraunhofer Institute for Integrated Circuits IIS, Erlangen \\
${^{\mathsection}}$ \textit{Multimedia Communications and Signal Processing}, Friedrich-Alexander-Universität Erlangen-Nürnberg \\
${^{\P}}$ \textit{Media: Conception and Production}, RheinMain University of Applied Sciences, Wiesbaden \\
Germany
}

\begin{document}
%
\maketitle
\begin{abstract}
In this work, we present a comparison between color spaces namely YUV, LAB, RGB and their effect on learned image compression. For this we use the structure and color based learned image codec (SLIC) from our prior work, which consists of two branches - one for the luminance component (Y or L) and another for chrominance components (UV or AB). However, for the RGB variant we input all 3 channels in a single branch, similar to most learned image codecs operating in RGB. The models are trained for multiple bitrate configurations in each color space. We report the findings from our experiments by evaluating them on various datasets and compare the results to state-of-the-art image codecs. The YUV model performs better than the LAB variant in terms of MS-SSIM with a Bj{\o}ntegaard delta bitrate (BD-BR) gain of 7.5\% using VTM intra-coding mode as the baseline. Whereas the LAB variant has a better performance than YUV model in terms of CIEDE2000 having a BD-BR gain of 8\%. Overall, the RGB variant of SLIC achieves the best performance with a BD-BR gain of  13.14\% in terms of MS-SSIM and a gain of 17.96\% in CIEDE2000 at the cost of a higher model complexity.	
	
\end{abstract}
\section{Introduction}\label{sec:intro}\thispagestyle{FirstPage}

One of the widely researched topics today is the use of deep neural networks in almost every field, with innumerable applications. In image processing and computer vision, they have been in vogue for many years. Recently, there has been a growing interest in the development of learned image codecs. A learned image codec uses non-linear neural networks consisting of several layers. Traditional image codecs such as JPEG  \cite{125072} use orthogonal linear mapping functions with discrete cosine transform (DCT). Learned image codecs are catching up with their traditional counterparts such as HEVC\cite{6316136} and  state-of-the-art VVC \cite{9503377} in the intra-coding mode.

Typically, learned image codecs are trained end-to-end in order to learn the encoder and decoder parameters jointly. The encoding process involves the transformation of an image into a latent representation, quantization and entropy coding. The quantized latent is extracted by the entropy decoder and reconstructed as an image by the non-linear decoder. This non-linear transform coding introduced in \cite{balleend} forms the basis for most learning based image codecs. The rate-distortion optimization (RDO) for such a system can be written as:
\begin{equation}
	\mathrm{min}_{\boldsymbol{\theta}, \boldsymbol{\phi}}\{L\}, \text{with  } L(\boldsymbol{\theta}, \boldsymbol{\phi}) = {R}({\boldsymbol{\theta}})   + \lambda \cdot D({\boldsymbol{\theta}, \boldsymbol{\phi}}),
	\label{eqn_rdo}
\end{equation} \\
where $L$ is the loss term,  $R$ is the rate measured in bits per pixel (bpp), $D$ represents distortion and $\lambda$ is the Lagrangian multiplier. The learnable parameters of the encoding and decoding networks are indicated by $\boldsymbol{\theta}$ and $\boldsymbol{\phi}$ respectively. 

A large part of learned image codecs operate in RGB color space. One of the reasons for this is the availability of image datasets in this color space. However, properties of the human visual system are not well exploited. Although the use of YUV color space in image compression is not new, in this work, we build on our prior works \cite{10222731, 10533739} to shed some light on the effect of color spaces in learned image compression. The performance is compared with state-of-the-art image codecs by means of rate-distortion curves, Bj{\o}ntegaard delta bitrate \cite{bjontegaard2001calculation}, and distortion values. Although we consider only RGB, YUV, and LAB, according to our knowledge, it is the first work in the domain of learned image compression focused on color spaces.

\begin{figure*}[th]
	\centering
	\begin{subfigure}[b]{0.45\textwidth}
		\includegraphics[width=1.0\textwidth]{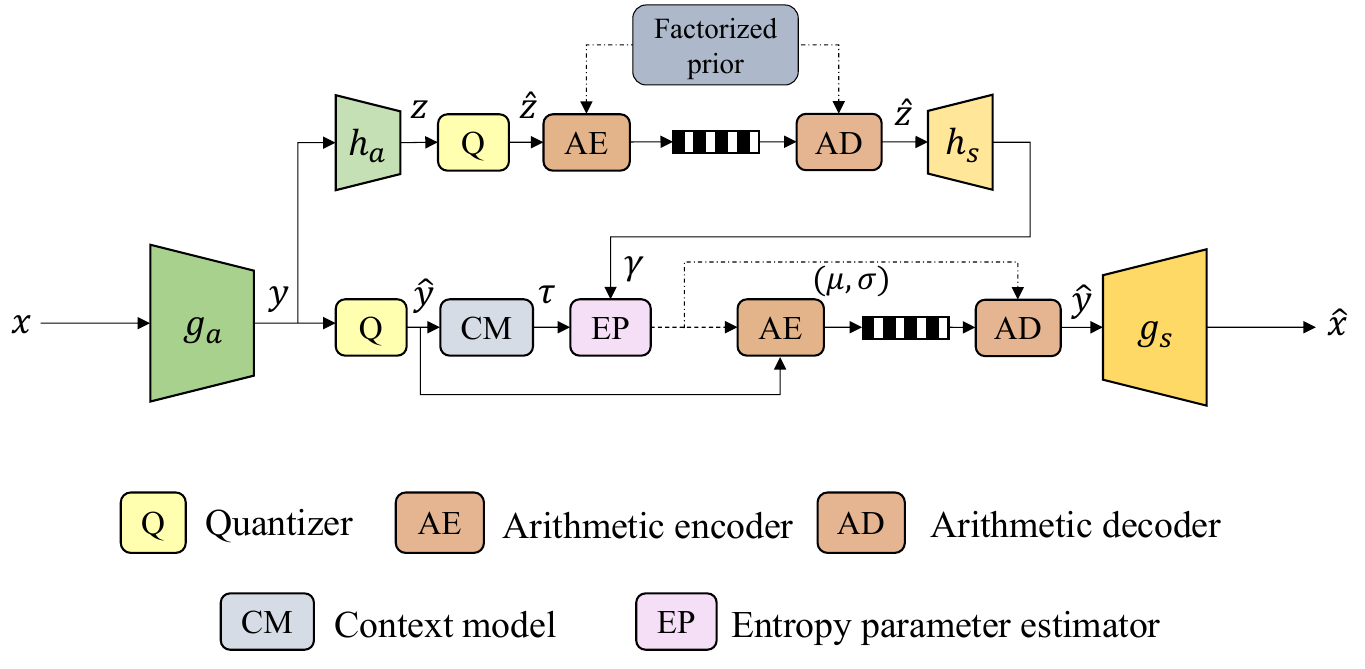} 
		\caption{SLIC--RGB}
		\label{fig:BlockRGB}
	\end{subfigure}
	\hfill
	\begin{subfigure}[b]{0.5\textwidth}
		\includegraphics[width=1.0\textwidth]{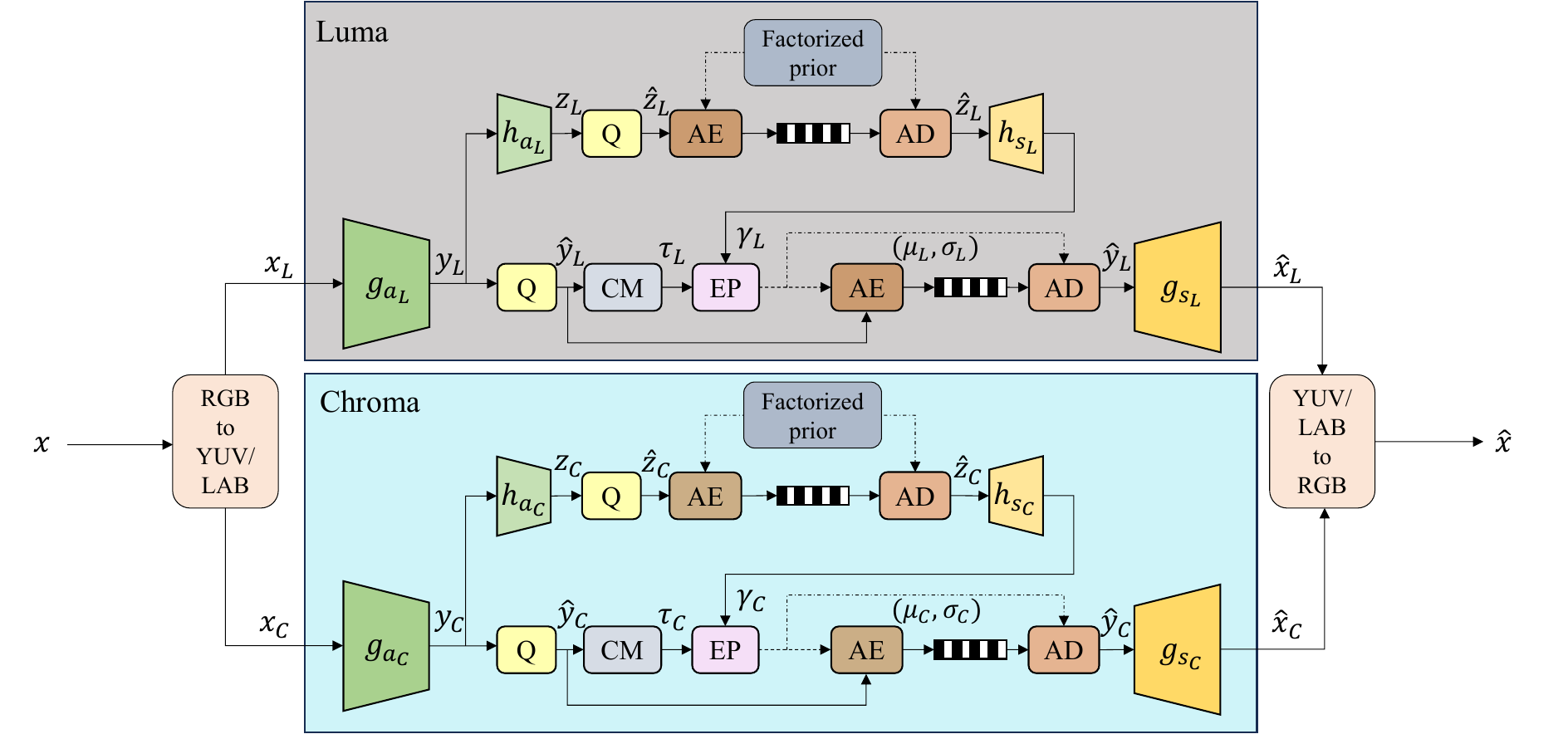}  
		\caption{SLIC--YUV / SLIC--LAB}
		\label{fig:BlockYUV}
	\end{subfigure}
	\caption{Network architecture of SLIC models.}
	\label{fig:BlockDiagram}
\end{figure*}
\section{Related work}\label{sec:sota}

Numerous contributions addressing specific problems in learned image compression have been made. The base framework for a majority of the learned codecs is variational autoencoder based. In order to have better entropy coding, a hyperprior model for forward adaptation is introduced in \cite{balle2018variational} and eventually extended with a backward adaptation in \cite{mbt2018}. The \emph{Cheng2020} \cite{cheng_learned_2020} was one of the first works to have competitive performance and the \emph{ELIC} introduced in \cite{he2022elic} outperforms VVC intra-coding mode \cite{9503377}.

The models in \cite{balle2018variational, mbt2018, cheng_learned_2020, he2022elic} are all trained and operate in RGB color space. There are some works such as \cite{10096867, 10018070} that work in the YUV color space. Learned image coding is also being made practical with standardization activities in JPEG AI \cite{10123093}, which aims to address several image processing and computer vision tasks in addition to compression. 

The effect of color space has been studied in general image classification \cite{gowda2019colornet} and robustness of deep learning \cite{de2021impact} to name a few. In our prior work \cite{10222731} and \cite{10533739}, we present our findings on designing a learning based codec that splits the task of image compression into two - structure from luminance channel and color from chrominance channels. It shows the advantage of optimizing networks with a color difference metric in addition to the other terms in the loss function.

\section{Color spaces in learned image compression}\label{sec:core}
Here we provide an overview of the SLIC model and its variants, the model architecture, workflow, loss function, and the implementation details including the training methodology.

\begin{figure*}[!th]
	\centering 
	\resizebox{0.625\textwidth}{!}{
\begin{tikzpicture}

\definecolor{darkgray176}{RGB}{176,176,176}
\definecolor{darkorange25512714}{RGB}{255,127,14}
\definecolor{forestgreen4416044}{RGB}{44,160,44}
\definecolor{goldenrod1911910}{RGB}{191,191,0}
\definecolor{green01270}{RGB}{0,127,0}
\definecolor{lightgray204}{RGB}{204,204,204}
\definecolor{steelblue31119180}{RGB}{31,119,180}
\tikzstyle{every node}=[font=\Large]

\begin{groupplot}[group style={group size=2 by 1, horizontal sep=1.75cm}]
\nextgroupplot[
tick align=outside,
tick pos=left,
x grid style={darkgray176},
x label style={at={(axis description cs:0.5,-0.025)},anchor=north},
xlabel={Rate [bpp]},
xmajorgrids,
xmin=0.0, xmax=1.0,
xtick style={color=black},
y grid style={darkgray176},
ylabel={PSNR [dB]   \(\displaystyle \uparrow\)},
ymajorgrids,
ymin=26.0, ymax=38.0,
ytick style={color=black}
]
\addplot [steelblue31119180, dash pattern=on 3.7pt off 1.6pt, mark=pentagon*, mark size=3, mark options={solid}]
table {%
0.098115338 27.98974522
0.173912113 29.82447267
0.299725056 31.8515377
0.456783226 33.87737664
0.679351191 35.62697935
0.829 36.654
};
\addplot [goldenrod1911910, dash pattern=on 3.7pt off 1.6pt, mark=triangle*, mark size=3, mark options={solid}]
table {%
0.132636176215278 27.5610487083333
0.210235595703125 29.1866699166667
0.321200900607639 30.931298
0.479699028862847 32.7869397083333
0.67002699110243 34.4550285
0.940165201822917 36.6320425833333
};
\addplot [red, dash pattern=on 3.7pt off 1.6pt, mark=triangle*, mark size=3, mark options={solid,rotate=180}]
table {%
0.123294406467014 26.8909755833333
0.189198811848958 28.1997195
0.288479275173611 29.5900013333333
0.441050211588542 31.2395902083333
0.6488037109375 32.9056854583333
0.967647976345486 35.3012495416667
};

\addplot [darkorange25512714, dash pattern=on 3.7pt off 1.6pt, mark=diamond, mark size=3, mark options={solid}]
table {%
0.1235 29.113
0.1965 30.698
0.3337 32.784
0.4908 34.577
0.7041 36.488
0.8573 37.624
};

\addplot [blue, mark=diamond*, mark size=3, mark options={solid}]
table {%
0.757168240017361 35.63334375
0.500481499565972 33.7509727083333
0.292887369791667 31.3773817083333
0.0870564778645833 27.6625960416667
};

\nextgroupplot[
legend cell align={left},
legend style={
  fill opacity=0.8,
  draw opacity=1,
  text opacity=1,
  at={(1.02,0.71)},
  anchor=west,
  draw=lightgray204
},
tick align=outside,
tick pos=left,
x grid style={darkgray176},
x label style={at={(axis description cs:0.5,-0.025)},anchor=north},
xlabel={Rate [bpp]},
xmajorgrids,
xmin=0.0, xmax=1.0,
xtick style={color=black},
y grid style={darkgray176},
ylabel={MS-SSIM [dB]   \(\displaystyle \uparrow\)},
ymajorgrids,
ymin=8.0, ymax=21.0,
ytick style={color=black}
]
\addplot [steelblue31119180, dash pattern=on 3.7pt off 1.6pt, mark=pentagon*, mark size=3, mark options={solid}]
table {%
0.098115338 10.678
0.173912113 12.36
0.299725056 14.567
0.456783226 16.476
0.679351191 18.201
0.829 19.166
};
\addlegendentry{Cheng2020 \cite{cheng_learned_2020}}
\addplot [goldenrod1911910, dash pattern=on 3.7pt off 1.6pt, mark=triangle*, mark size=3, mark options={solid}]
table {%
0.132636176215278 10.8233301010038
0.210235595703125 12.3639873816379
0.321200900607639 14.0854223288858
0.479699028862847 15.9817091981016
0.67002699110243 17.81696758263
0.940165201822917 19.6487789593679
};
\addlegendentry{Hyper Prior \cite{balle2018variational}}
\addplot [red, dash pattern=on 3.7pt off 1.6pt, mark=triangle*, mark size=3, mark options={solid,rotate=180}]
table {%
0.123294406467014 10.4523575597074
0.189198811848958 11.934573295591
0.288479275173611 13.5046738838869
0.441050211588542 15.2670919407658
0.6488037109375 17.1102471554952
0.967647976345486 19.131352100209
};
\addlegendentry{Factorized Prior \cite{balleend}}

\addplot [darkorange25512714, dash pattern=on 3.7pt off 1.6pt, mark=diamond, mark size=3, mark options={solid}]
table {%
	0.8573 37.624
};
\addlegendentry{ELIC MSE \cite{he2022elic}}

\addplot [darkorange25512714, dash pattern=on 3.7pt off 1.6pt, mark=diamond*, mark size=3, mark options={solid}]
table {%
0.1242 11.593
0.197 13.258
0.3342 15.456
0.4916 17.272
0.7045 19.059
0.8574 20.116
};
\addlegendentry{ELIC MS-SSIM \cite{he2022elic}}

\addplot [blue, mark=diamond*, mark size=3, mark options={solid}]
table {%
0.757168240017361 19.381393678815
0.500481499565972 17.3330692433674
0.292887369791667 14.5476420978346
0.0870564778645833 10.5008009930969
};
\addlegendentry{SLIC--RGB (ours)}
\end{groupplot}

\end{tikzpicture}}
	\caption{RD curves of learned image codecs operating in RGB for the \emph{Kodak} dataset .}
	\label{fig:RD_all}
\end{figure*}
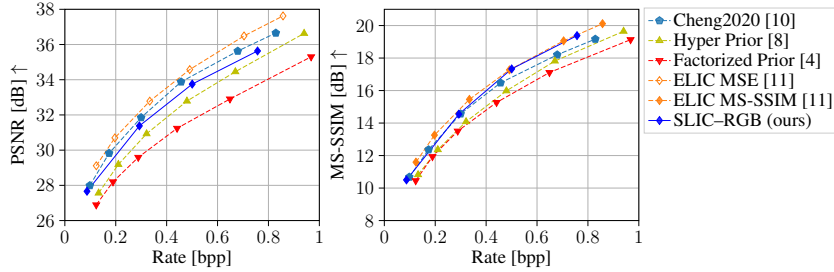

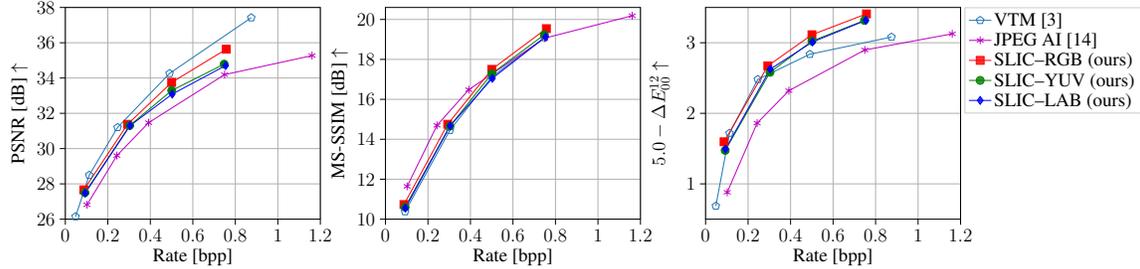
\begin{figure*}[!th]
	\centering 
	\resizebox{0.85\textwidth}{!}{
\begin{tikzpicture}

\definecolor{darkgray176}{RGB}{176,176,176}
\definecolor{steelblue31119180}{RGB}{31,119,180}
\definecolor{darkviolet1910191}{RGB}{191,0,191}
\definecolor{green01270}{RGB}{0,127,0}
\definecolor{lightgray204}{RGB}{204,204,204}
\tikzstyle{every node}=[font=\Large]

\begin{groupplot}[group style={group size=3 by 1, horizontal sep=1.75cm}]
\nextgroupplot[
tick align=outside,
tick pos=left,
x grid style={darkgray176},
xlabel={Rate [bpp]},
x label style={at={(axis description cs:0.5,-0.025)},anchor=north},
xmajorgrids,
xmin=0.0, xmax=1.2,
xtick style={color=black},
y grid style={darkgray176},
ylabel={PSNR [dB] \(\displaystyle \uparrow\)},
ymajorgrids,
ymin=26.0, ymax=38.0,
ytick style={color=black}
]
\addplot [line width=0.3pt, steelblue31119180, mark=pentagon, mark size=3, mark options={solid}]
table {%
0.0482449 26.14484539
0.112495422 28.49302175
0.245816549 31.19987372
0.490545485 34.26153257
0.874810113 37.41979316
};
\addplot [line width=0.3pt, darkviolet1910191, mark=asterisk, mark size=3, mark options={solid}]
table {%
0.101913452 26.81502479
0.242577447 29.60188679
0.392044915 31.46900596
0.7493337 34.19210021
1.160803901 35.27101025
};
\addplot [line width=0.3pt, red, mark=square*, mark size=3, mark options={solid}]
table {%
0.757168240017361 35.63334375
0.500481499565972 33.7509727083333
0.292887369791667 31.3773817083333
0.0870564778645833 27.6625960416667
};
\addplot [line width=0.3pt, green01270, mark=*, mark size=3, mark options={solid}]
table {%
0.747585720486111 34.78781825
0.500698513454861 33.2932780416667
0.303198920355903 31.3122765416667
0.0923055013020833 27.4999881666667
};
\addplot [line width=0.3pt, blue, mark=diamond*, mark size=3, mark options={solid}]
table {%
0.754174126519097 34.7043182916667
0.502987331814236 33.0877543333333
0.304365370008681 31.299487
0.0936347113715277 27.480973125
};

\nextgroupplot[
tick align=outside,
tick pos=left,
x grid style={darkgray176},
xlabel={Rate [bpp]},
x label style={at={(axis description cs:0.5,-0.025)},anchor=north},
xmajorgrids,
xmin=0.0, xmax=1.2,
xtick style={color=black},
y grid style={darkgray176},
ylabel={MS-SSIM [dB] \(\displaystyle \uparrow\)},
ymajorgrids,
ymin=10.00, ymax=20.6,
ytick style={color=black}
]
\addplot [line width=0.3pt, steelblue31119180, mark=pentagon, mark size=3, mark options={solid}]
table {%
0.747585720486111 19.0923095513288
0.500698513454861 17.1210685025709
0.303198920355903 14.4531300951154
0.0923055013020833 10.3767280919996
};
\addplot [line width=0.3pt, darkviolet1910191, mark=asterisk, mark size=3, mark options={solid}]
table {%
0.101913452 11.64061077
0.242577447 14.69673408
0.392044915 16.48289269
0.7493337 19.0632285
1.160803901 20.18566558
};
\addplot [line width=0.3pt, red, mark=square*, mark size=3, mark options={solid}]
table {%
0.757168240017361 19.5402807083333
0.500481499565972 17.5013136666667
0.292887369791667 14.7530575416667
0.0870564778645833 10.7307583125
};
\addplot [line width=0.3pt, green01270, mark=*, mark size=3, mark options={solid}]
table {%
0.747585720486111 19.2340878333333
0.500698513454861 17.2873269166667
0.303198920355903 14.6540652916667
0.0923055013020833 10.6018311875
};
\addplot [line width=0.3pt, blue, mark=diamond*, mark size=3, mark options={solid}]
table {%
0.754174126519097 19.12705975
0.502987331814236 17.0522122916667
0.304365370008681 14.673846375
0.0936347113715277 10.5551457916667
};

\nextgroupplot[
legend cell align={left},
legend style={
  fill opacity=0.8,
  draw opacity=1,
  text opacity=1,
  at={(1.71, 0.50)},
  anchor=south east,
  draw=lightgray204
},
tick align=outside,
tick pos=left,
x grid style={darkgray176},
x label style={at={(axis description cs:0.5,-0.025)},anchor=north},
xlabel={Rate [bpp]},
xmajorgrids,
xmin=0.0, xmax=1.2,
xtick style={color=black},
y grid style={darkgray176},
ylabel={$5.0 - \Delta E_{00}^{12}$ \(\displaystyle \uparrow\)},
ymajorgrids,
ymin=0.50, ymax=3.5,
ytick style={color=black}
]
\addplot  [line width=0.3pt, steelblue31119180, mark=pentagon, mark size=3, mark options={solid}]
table {%
0.0482449 0.686358079
0.112495422 1.718549475
0.245816549 2.480570932
0.490545485 2.837229937
0.874810113 3.080747209
};
\addlegendentry{VTM\cite{9503377}}
\addplot [line width=0.3pt, darkviolet1910191, mark=asterisk, mark size=3, mark options={solid}]
table {%
0.101913452 0.880422933
0.242577447 1.860981937
0.392044915 2.320655283
0.7493337 2.89755025
1.160803901 3.126500367
};
\addlegendentry{JPEG AI \cite{10123093}}
\addplot [line width=0.3pt, red, mark=square*, mark size=3, mark options={solid}]
table {%
0.757168240017361 3.40909435416667
0.500481499565972 3.11385335833333
0.292887369791667 2.67192866666667
0.0870564778645833 1.59603944583333
};
\addlegendentry{SLIC--RGB (ours)}
\addplot [line width=0.3pt, green01270, mark=*, mark size=3, mark options={solid}]
table {%
0.747585720486111 3.31574027916667
0.500698513454861 3.02383842083333
0.303198920355903 2.58113147083333
0.0923055013020833 1.47395596666667
};
\addlegendentry{SLIC--YUV (ours)}
\addplot [line width=0.3pt, blue, mark=diamond*, mark size=3, mark options={solid}]
table {%
0.754174126519097 3.31763234166667
0.502987331814236 3.01060720833333
0.304365370008681 2.62556040833333
0.0936347113715277 1.49415787083333
};
\addlegendentry{SLIC--LAB (ours)}
\end{groupplot}

\end{tikzpicture}}
	\caption{RD curves of learned image codecs operating in YUV (JPEG AI and SLIC--YUV), SLIC--LAB, SLIC--RGB, and VTM for the \emph{Kodak} dataset.}
	\label{fig:RD_yuv}
\end{figure*}

\subsection{Overview}
We build on the structure and color based learned image codec (SLIC) from our prior work in \cite{10533739}, which is an image codec operating in YUV. In this paper, we introduce two new variants of the SLIC model called SLIC--LAB and SLIC--RGB. But for clarity, the original SLIC model will henceforth be referred to as SLIC--YUV. SLIC--LAB has exactly the same architecture as the SLIC--YUV model except the operating color space. To have an equivalent RGB model with same set of layers, we introduce the SLIC--RGB, which has a single branch. The model architectures are shown in Fig.\ \ref{fig:BlockDiagram}. The SLIC--RGB model is as shown in Fig.\ \ref{fig:BlockRGB} and Fig.\ \ref{fig:BlockYUV} indicates the block diagram for the YUV or LAB variant, wherein the input and output images are in RGB, but the operating color space is either YUV or LAB. 
The internal details of the models are discussed in greater detail in our prior work \cite{10533739}. In the RGB model, the single branch consists of 192 channels. In case of YUV and LAB models, the luminance branch consists of 128 channels and the chrominance branch has 64 channels.  Due to the increase in the number of channels in a single branch, SLIC--RGB has a higher complexity. The two-branch models have around 20 million parameters and need 1,512 kilo multiply accumulate operations per pixel (kMACs/pixel) including encoder and decoder. For the RGB variant, it is about 36 million parameters with 4,840 kMACs/pixel.

In the SLIC--RGB model in Fig.\ \ref{fig:BlockRGB}, the analysis transform $g_a$ translates the image $x$ into a latent space representation $y$. This is further transformed, to learn the statistical distribution of the latent as a hyperlatent representation $z$ through the hyper analysis transform $h_a$. A factorized prior model helps in encoding the quantized hyperlatents $\hat{z}$. The quantized latent $\hat{y}$ is efficiently encoded by backward adaptation through an autoregressive context model (CM) using masked convolution as proposed in \cite{mbt2018}. The entropy parameter estimation module (EP) combines the output $\gamma$ of hypersynthesis transform $h_s$, and $\tau$ from the context model, to predict mean $\mu$ and variance $\sigma$. In other words, the probability distribution of the latent is estimated and used for entropy coding. The entropy decoded latent $\hat{y}$ is then transformed back to image space through the synthesis transform $g_s$ as image $\hat{x}$.

For the YUV and LAB models shown in Fig.\ \ref{fig:BlockYUV}, the same workflow holds true. However, an input image is converted from RGB into the respective color space and split into luminance and chrominance components. These are then fed into their corresponding branches. The symbols with subscripts $\cdot_L$ and $\cdot_C$ indicate luminance and chrominance components respectively. The outputs of these networks are combined and converted back to an RGB image.
\subsection{Loss Function} 
RDO is the backbone in optimizing image codecs. In this work, we use an objective function that is a combination of three distortion metrics and the rate term. The metrics used are mean squared error (MSE), multi-scale structural similarity index measure (MS-SSIM) \cite{wang_multiscale_2003}, and color difference metric CIEDE2000 ($\Delta E_{00}^{12}$) \cite{sharma2005ciede2000}. The loss function, as used in our prior works \cite{10222731, 10533739} is :
\begin{equation}
	\begin{aligned}
		\mathrm{min}_{\boldsymbol{\theta}, \boldsymbol{\phi}}\{L\}, \text{with  } L(\boldsymbol{\theta}, \boldsymbol{\phi}) = {R}  + \lambda_{1} \cdot \mathrm{MSE}(x, \hat{x}) \\
		+ \lambda_{2} \cdot (1.0 - \mathrm{MS\text{-}SSIM}(x, \hat{x}))
		+   \lambda_{3} \cdot \Delta E_{00}^{12}(x, \hat{x}),
		\label{eqn:loss}
	\end{aligned}
\end{equation}
where $\lambda_{1}, \lambda_{2}$, and $\lambda_{3}$ are the Lagrangian multipliers for the metrics MSE, MS-SSIM and CIEDE2000 respectively. It should be noted that MSE and MS-SSIM are estimated in RGB color space, since the data used for training and evaluating the models are RGB images. All the SLIC variants are trained with this same loss function.

The rate term $R$ comprises of the bits required to encode the image $x$. For the RGB model, it constitues of the latent bits and the hyperlatent bits. However, in YUV or LAB model, there are luma and chroma branches, contributing a total of four components.

\subsection{Implementation details} \label{ssec:imp}
The models are implemented in Python programming language using the \texttt{PyTorch}\footnote[1]{\url{https://pytorch.org}} framework and \texttt{CompressAI} library \cite{begaint2020compressai}.

The model variants are trained individually for each operating color space, with the loss function in (\ref{eqn:loss}) and four operating points with Lagrangian values from our prior works \cite{10222731, 10533739}; $\lambda_{1}=\{0.001, 0.005, 0.01, 0.02\}$, $\lambda_{2}=\{0.01, 0.12, 2.4, 4.8\}$, and $\lambda_{3}=\{0.024, 0.12, 0.24, 0.48\}$ for MSE, MS-SSIM and $ \Delta E_{00}^{12}$ respectively. The models are trained for \textrm{120} epochs with the \emph{COCO2017} training dataset \cite{lin2014microsoft} comprising around \textrm{118,000} images. The validation data has about $\textrm{5,000}$ randomly chosen images from the \emph{ImageNet} dataset \cite{5206848} spanning various classes. Adam optimizer \cite{kingma2017adam} initialized with a learning rate of $\textrm{1e-4}$ is employed in tandem with a learning rate scheduler.
\section{Experiments and results}\label{sec:experiments}
In this section, we report our findings from various experiments. We start with the rate-distortion performance where we compare various codecs using the metrics PSNR, MS-SSIM and CIEDE2000 for model configurations resulting in a bitrate range of 0 to 1 bpp. Followed by that, we make a comparison between the effect of color channels in the SLIC--LAB and SLIC--YUV models. Finally, we illustrate the effect of color spaces on the latent channels through the channel impulse responses and visual comparison. 

\begin{table*}[!th]
	\centering
	\small
	\caption{BD-Rate and BD-Distortion values of various image codecs with VTM-intra baseline for the \emph{Kodak} dataset.}
	\label{tab:bd_rate}
	\vspace{-3pt}
	\begin{tabular}{|c|cc|cc|cc|}
		\hline
		\multirow{2}{*}{Codec Name} & \multicolumn{2}{c|}{PSNR}        & \multicolumn{2}{c|}{MS-SSIM}        & \multicolumn{2}{c|}{CIEDE2000}         \\ \cline{2-7} 
		& \multicolumn{1}{c|}{\begin{tabular}[c]{@{}c@{}}BD-BR\\ (\%) \end{tabular}} & \begin{tabular}[c]{@{}c@{}}BD-PSNR \\ (dB) \end{tabular} & \multicolumn{1}{c|}{\begin{tabular}[c]{@{}c@{}}BD-BR\\ (\%) \end{tabular}} & \begin{tabular}[c]{@{}c@{}}BD-MS-\\SSIM\end{tabular}  & \multicolumn{1}{c|}{\begin{tabular}[c]{@{}c@{}}BD-BR\\ (\%) \end{tabular}} & \begin{tabular}[c]{@{}c@{}}BD-1\\/CIEDE2000\end{tabular} \\ \hline
		Cheng2020  \cite{cheng_learned_2020}  & \multicolumn{1}{c|}{\underline{3.40}}   & \underline{-0.1461}   & \multicolumn{1}{c|}{{-3.32}}  &{ {0.1333}}   & \multicolumn{1}{c|}{{20.82}}   & {-0.0175} \\
		ELIC MSE \cite{he2022elic}              & \multicolumn{1}{c|}{{\textbf{-7.07}}}   & {\textbf{0.3260}}   & \multicolumn{1}{c|}{{--}}  & {--}   & \multicolumn{1}{c|}{{--}}   & {--} \\
		ELIC MS-SSIM \cite{he2022elic}              & \multicolumn{1}{c|}{{{--}}}   & {{--}}   & \multicolumn{1}{c|}{{-12.87}}  & \underline{0.5961}   & \multicolumn{1}{c|}{{--}}   & {--} \\
		JPEG AI\cite{10123093}  & \multicolumn{1}{c|}{{{55.75}}}   & {{-1.7562}}   & \multicolumn{1}{c|}{\textbf{-20.16}}  & \textbf{0.9121}   & \multicolumn{1}{c|}{{69.68}}   & {-0.6138} \\
		SLIC--RGB (Ours)                  & \multicolumn{1}{c|}{{{12.60}}}   & {{-0.5298}}   & \multicolumn{1}{c|}{\underline{-13.14}}  & {0.4772}   & \multicolumn{1}{c|}{\textbf{-17.96}}   & \textbf{0.0302} \\
		SLIC--YUV (Ours)                  & \multicolumn{1}{c|}{{21.73}}   & {-0.8274}   & \multicolumn{1}{c|}{{{-7.50}}}  & {0.2052}   & \multicolumn{1}{c|}{{-4.66}}   & {0.0080} \\
		SLIC--LAB (Ours)                  & \multicolumn{1}{c|}{{22.65}}   & {-0.8305}   & \multicolumn{1}{c|}{{{-6.23}}}  & {0.2342}   & \multicolumn{1}{c|}{\underline{-7.99}}   & \underline{0.0157} \\
		\hline
	\end{tabular}
	\vspace{2pt}
	\\ \textbf{Bold} indicates the best values and \underline{underline} represents the second best.
\end{table*}

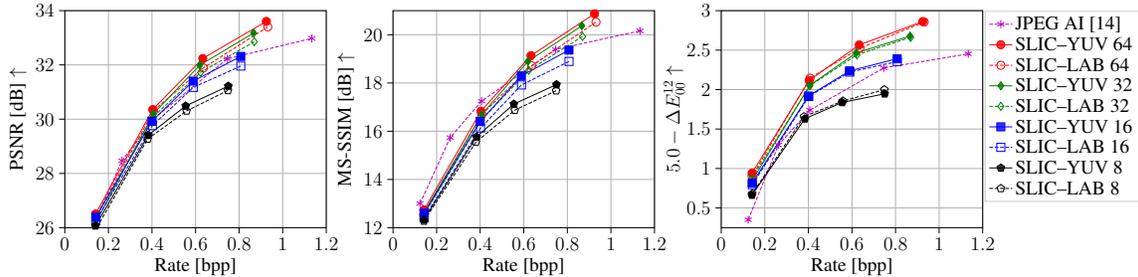
\begin{figure*}[th]
	\centering
	\resizebox{0.85\textwidth}{!}{
\begin{tikzpicture}

\definecolor{darkgray176}{RGB}{176,176,176}
\definecolor{darkviolet1910191}{RGB}{191,0,191}
\definecolor{green01270}{RGB}{0,127,0}
\definecolor{lightgray204}{RGB}{204,204,204}
\tikzstyle{every node}=[font=\Large]

\begin{groupplot}[group style={group size=3 by 1, horizontal sep=1.75cm}]
\nextgroupplot[
tick align=outside,
tick pos=left,
x grid style={darkgray176},
x label style={at={(axis description cs:0.5,-0.025)},anchor=north},
xlabel={Rate [bpp]},
xmajorgrids,
xmin=0.0, xmax=1.2,
xtick style={color=black},
y grid style={darkgray176},
ylabel={PSNR [dB] \(\displaystyle \uparrow\)},
ymajorgrids,
ymin=26.0, ymax=34.0,
ytick style={color=black}
]
\addplot [line width=0.3pt, darkviolet1910191, dash pattern=on 2.775pt off 1.2pt, mark=asterisk, mark size=3, mark options={solid}]
table {%
0.123808 25.93944515
0.262236 28.46711336
0.406316 30.00843659
0.746041 32.22455913
1.1336755 32.97733429
};
\addplot [line width=0.3pt, red, mark=*, mark size=3, mark options={solid}]
table {%
0.924814 33.6026942
0.633028 32.23700357
0.403472 30.35658852
0.141496 26.50566422
};
\addplot [line width=0.3pt, red, dash pattern=on 2.775pt off 1.2pt, mark=o, mark size=3, mark options={solid}]
table {%
0.93163 33.40223496
0.636134 31.89053317
0.407426 30.31331882
0.141476 26.37437881
};
\addplot [line width=0.3pt, green01270, mark=diamond*, mark size=3, mark options={solid}]
table {%
0.866468 33.17557548
0.619402 31.99237473
0.406718 30.19828579
0.140944 26.16930459
};
\addplot [line width=0.3pt, green01270, dash pattern=on 2.775pt off 1.2pt, mark=diamond, mark size=3, mark options={solid}]
table {%
0.869658 32.85605919
0.622894 31.72557756
0.40601 30.08468711
0.140396 26.3538119
};
\addplot [line width=0.3pt, blue, mark=square*, mark size=3, mark options={solid}]
table {%
0.807412 32.31321591
0.58896 31.40228316
0.399776 29.92437209
0.142266 26.38782919
};
\addplot [line width=0.3pt, blue, dash pattern=on 2.775pt off 1.2pt, mark=square, mark size=3, mark options={solid}]
table {%
0.807774 31.95685919
0.589712 31.17127473
0.4019 29.7632258
0.14257 26.2559767
};
\addplot [line width=0.3pt, black, mark=pentagon*, mark size=3, mark options={solid}]
table {%
0.750176 31.21936692
0.554606 30.48707601
0.38323 29.41120272
0.14065 26.06670884
};
\addplot [line width=0.3pt, black, dash pattern=on 2.775pt off 1.2pt, mark=pentagon, mark size=3, mark options={solid}]
table {%
0.748638 31.05928063
0.558796 30.3025657
0.380166 29.26891677
0.14191 25.9359111
};

\nextgroupplot[
tick align=outside,
tick pos=left,
x grid style={darkgray176},
xlabel={Rate [bpp]},
x label style={at={(axis description cs:0.5,-0.025)},anchor=north},
xmajorgrids,
xmin=0.0, xmax=1.2,
xtick style={color=black},
y grid style={darkgray176},
ylabel={MS-SSIM [dB] \(\displaystyle \uparrow\)},
ymajorgrids,
ymin=12, ymax=21,
ytick style={color=black}
]
\addplot [line width=0.3pt, darkviolet1910191, dash pattern=on 2.775pt off 1.2pt, mark=asterisk, mark size=3, mark options={solid}]
table {%
0.123808 13.00139518
0.262236 15.73545391
0.406316 17.24988458
0.746041 19.39163631
1.1336755 20.16062475
};
\addplot [line width=0.3pt, red, mark=*, mark size=3, mark options={solid}]
table {%
0.924814 20.87045523
0.633028 19.13510349
0.403472 16.832795555
0.141496 12.733043975
};
\addplot [line width=0.3pt, red, dash pattern=on 2.775pt off 1.2pt, mark=o, mark size=3, mark options={solid}]
table {%
0.93163 20.52517851
0.636134 18.71582214
0.407426 16.69173672
0.141476 12.549294665
};
\addplot [line width=0.3pt, green01270, mark=diamond*, mark size=3, mark options={solid}]
table {%
0.866468 20.37164165
0.619402 18.90072839
0.406718 16.685846635
0.140944 12.614733395
};
\addplot [line width=0.3pt, green01270, dash pattern=on 2.775pt off 1.2pt, mark=diamond, mark size=3, mark options={solid}]
table {%
0.869658 19.92977074
0.622894 18.52766316
0.40601 16.46666725
0.140396 12.51090661
};
\addplot [line width=0.3pt, blue, mark=square*, mark size=3, mark options={solid}]
table {%
0.807412 19.36497758
0.58896 18.293469165
0.399776 16.40664374
0.142266 12.60317641
};
\addplot [line width=0.3pt, blue, dash pattern=on 2.775pt off 1.2pt, mark=square, mark size=3, mark options={solid}]
table {%
0.807774 18.89485378
0.589712 17.918780205
0.4019 16.132316095
0.14257 12.409314025
};
\addplot [line width=0.3pt, black, mark=pentagon*, mark size=3, mark options={solid}]
table {%
0.750176 17.933109195
0.554606 17.127458655
0.38323 15.766912665
0.14065 12.32277751
};
\addplot [line width=0.3pt, black, dash pattern=on 2.775pt off 1.2pt, mark=pentagon, mark size=3, mark options={solid}]
table {%
0.748638 17.67921197
0.558796 16.87577189
0.380166 15.54686408
0.14191 12.27467224
};

\nextgroupplot[
legend cell align={left},
legend style={
  fill opacity=0.8,
  draw opacity=1,
  text opacity=1,
  at={(1.60,0.12)},
  anchor=south east,
  draw=lightgray204
},
tick align=outside,
tick pos=left,
x grid style={darkgray176},
xlabel={Rate [bpp]},
x label style={at={(axis description cs:0.5,-0.025)},anchor=north},
xmajorgrids,
xmin=0.0, xmax=1.2,
xtick style={color=black},
y grid style={darkgray176},
ylabel={$5.0 - \Delta E_{00}^{12}$ \(\displaystyle \uparrow\)},
ymajorgrids,
ymin=0.25, ymax=3.0,
ytick style={color=black}
]
\addplot [line width=0.3pt, darkviolet1910191, dash pattern=on 2.775pt off 1.2pt, mark=asterisk, mark size=3, mark options={solid}]
table {%
0.123808 0.350073687
0.262236 1.297214114
0.406316 1.739342048
0.746041 2.273648704
1.1336755 2.452217498
};
\addlegendentry{JPEG AI \cite{10123093}}
\addplot [line width=0.3pt, red, mark=*, mark size=3, mark options={solid}]
table {%
0.924814 2.8619477354
0.633028 2.5659123158
0.403472 2.115936268
0.141496 0.942174046
};
\addlegendentry{SLIC--YUV 64}
\addplot [line width=0.3pt, red, dash pattern=on 2.775pt off 1.2pt, mark=o, mark size=3, mark options={solid}]
table {%
0.93163 2.8567552905
0.636134 2.5233094173
0.407426 2.1448470671
0.141476 0.902292129
};
\addlegendentry{SLIC--LAB 64}
\addplot [line width=0.3pt, green01270, mark=diamond*, mark size=3, mark options={solid}]
table {%
0.866468 2.6776934085
0.619402 2.464167491
0.406718 2.0544628519
0.140944 0.841819279000001
};
\addlegendentry{SLIC--YUV 32}
\addplot [line width=0.3pt, green01270, dash pattern=on 2.775pt off 1.2pt, mark=diamond, mark size=3, mark options={solid}]
table {%
0.869658 2.6621807996
0.622894 2.4402819468
0.40601 2.0469381175
0.140396 0.873253243000001
};
\addlegendentry{SLIC--LAB 32}
\addplot [line width=0.3pt, blue, mark=square*, mark size=3, mark options={solid}]
table {%
0.807412 2.3907327863
0.58896 2.236484813
0.399776 1.917014221
0.142266 0.815162881000001
};
\addlegendentry{SLIC--YUV 16}
\addplot [line width=0.3pt, blue, dash pattern=on 2.775pt off 1.2pt, mark=square, mark size=3, mark options={solid}]
table {%
0.807774 2.3552445864
0.589712 2.2212937116
0.4019 1.912862327
0.14257 0.794682572
};
\addlegendentry{SLIC--LAB 16}
\addplot [line width=0.3pt, black, mark=pentagon*, mark size=3, mark options={solid}]
table {%
0.750176 1.946109444
0.554606 1.836580097
0.38323 1.62884079
0.14065 0.662340876999999
};
\addlegendentry{SLIC--YUV 8}
\addplot [line width=0.3pt, black, dash pattern=on 2.775pt off 1.2pt, mark=pentagon, mark size=3, mark options={solid}]
table {%
0.748638 1.9981362348
0.558796 1.855692949
0.380166 1.657288904
0.14191 0.680143317
};
\addlegendentry{SLIC--LAB 8}
\end{groupplot}

\end{tikzpicture}}
	\caption{RD curves of chroma channel variants of SLIC--LAB and SLIC--YUV models for the \emph{Tecknick} RGB dataset.}
	\label{fig:RD_channels}
\end{figure*}

\subsection{Rate-distortion performance}
We measure rate-distortion (RD) performance of various codecs and compare them with our models using the \emph{Kodak} dataset, that consists of 24 images of resolution $512\times768$ in either orientation. This experiment is split into two parts. First, we compare the RGB codecs and then the codecs operating in YUV. In both cases, we compute RGB PSNR and MS-SSIM with the original and reconstructed images for all bitrate configurations for each codec. 

The PSNR for each RGB image is computed as the average across each pixel over every channel. Similarly, the MS-SSIM metric is calculated according to \cite{wang_multiscale_2003} and by using equal weights for RGB channels. 

The CIEDE2000 metric, indicated by $\Delta E_{00}^{12}$ requires a color conversion from RGB to LAB. Moreover, we represent the metric as $5.0 - \Delta E_{00}^{12}$ in order to complement it as a quality metric, with $5.0$ as an offset based on the range of values.

\begin{figure}[!thb]
	\centering
	\subfloat[\centering SLIC--YUV \newline Rate: $0.4398$ bpp, PSNR-Luma: $37.23$ dB \newline]{{\includegraphics[height=0.28\textwidth]{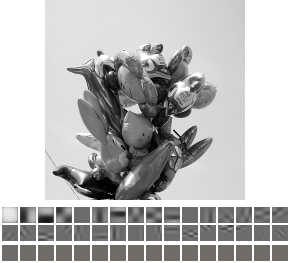}}}
	\qquad
	\subfloat[\centering SLIC--LAB \newline Rate: $0.4410$ bpp, PSNR-Luma: $37.31$ dB \newline]{{\includegraphics[height=0.28\textwidth]{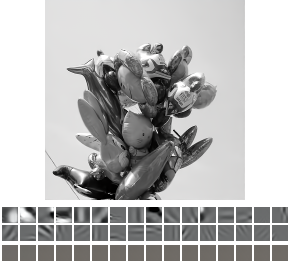} }}
	\qquad
	\subfloat[\centering SLIC--RGB \newline Rate: $0.4657$ bpp, PSNR-Luma: $37.90$ dB \newline ]{{\includegraphics[height=0.28\textwidth]{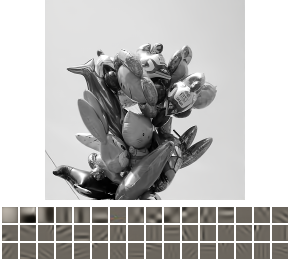} }}
	\caption{\centering Reconstructed versions and channel impulse responses of the image \texttt{GRAY\_R03\_0400x0400\_014.png}}%
	\label{fig:Gray}
\end{figure}

\begin{figure}[!ht]
	\centering
	\subfloat[\centering SLIC--YUV \newline Rate: $0.7158$ bpp, PSNR: $34.49$ dB, \newline MS-SSIM: $22.29$ dB, $\Delta E_{00}$ :  $1.56$]{{\includegraphics[height=0.28\textwidth]{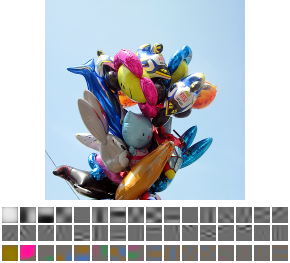} }}
	\qquad
	\subfloat[\centering SLIC--LAB \newline Rate: $0.7205$ bpp, PSNR: $33.90$ dB, \newline MS-SSIM: $21.62$ dB, $\Delta E_{00}$ :  $1.56$]{{\includegraphics[height=0.28\textwidth]{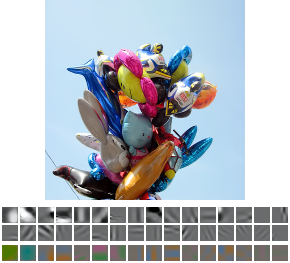} }}
	\qquad
	\subfloat[\centering SLIC--RGB \newline Rate: $0.6379$ bpp, PSNR: $35.08$ dB, \newline MS-SSIM: $22.75$ dB, $\Delta E_{00}$ :  $1.47$]{{\includegraphics[height=0.28\textwidth]{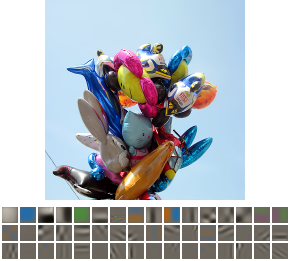} }}
	\caption{\centering Reconstructed versions and channel impulse responses of the image \texttt{RGB\_R03\_0400x0400\_014.png}}%
	\label{fig:RGB}%
\end{figure}

\subsubsection{RGB image codecs}
A comparison of rate-distortion performance of SLIC--RGB is made with \textit{Cheng2020}\cite{cheng_learned_2020}, \textit{ELIC}\cite{he2022elic}, \textit{Hyper Prior}\cite{balle2018variational}, and \textit{Factorized Prior}\cite{balleend} models. The rate-distortion curves are reported in Fig.\ \ref{fig:RD_all} for PSNR and MS-SSIM. We use the RD values of ELIC from CompressAI, where PSNR values are provided for MSE optimized ELIC and MS-SSIM values for MS-SSIM optimized models. They are indicated as ELIC--MSE and ELIC--MS-SSIM respectively. 

In terms of PSNR, ELIC--MSE has the best performance and Cheng2020 is better than SLIC--RGB. When we observe the MS-SSIM curves, ELIC--MS-SSIM has the best performance. However, SLIC--RGB outperforms Cheng2020, and is comparable to ELIC at bitrates larger than 0.5 bpp.

\subsubsection{YUV image codecs}
We compare \textit{JPEG AI }\cite{10123093} and  intra-coding mode of VVC test model (VTM)\cite{9503377} with the SLIC variants. The RD curves are presented in Fig.\ \ref{fig:RD_yuv} with PSNR, MS-SSIM and $\Delta E_{00}^{12}$ metrics at various bitrates. For JPEG AI, the verification model \texttt{vm-release-v4.5} is used in the default evaluation mode. Clearly, VTM has the best PSNR performance. With MS-SSIM, JPEG AI outperforms all codecs under consideration for bitrates lower than 0.5 bpp. But VTM, SLIC--LAB, and SLIC--YUV variants are close in terms of MS-SSIM and SLIC--RGB has slightly higher values.

JPEG AI \cite{10123093} has the worst performance in terms of CIEDE2000. VTM has a comparable performance to SLIC--RGB for bitrates lower than 0.2 bpp. It is interesting to see that SLIC variants have a superior performance and thus, having a color difference metric in the loss function can improve color fidelity.

\subsubsection{BD-Rate and BD-Distortion}
The Bj{\o}ntegaard-delta bitrate \cite{bjontegaard2001calculation} and distortion values are measured and reported in Table\ \ref{tab:bd_rate}. VTM is used as the baseline. The values are measured for PSNR, MS-SSIM, and CIEDE2000 metrics. In each column, the best value is indicated in bold and the second best value is underlined. The MSE optimized ELIC model has the most gains in PSNR, both in bitrate and distortion with values $7.07\%$ and $0.326$ dB respectively. In the MS-SSIM column, SLIC--RGB has a gain of $13.14\%$ in BD-BR. However, JPEG AI has the highest gain with $20.16\%$ and $0.9121$ dB in bitrate and distortion respectively. 

Lastly, for the color difference metric CIEDE2000, only the SLIC variants have a BD-BR gain of $17.96\%$ for SLIC--RGB, $7.99\%$ for SLIC--LAB, and a gain of $4.66\%$ for SLIC--YUV models. This is also reflected in the RD curves in Fig.\ \ref{fig:RD_yuv}. With the color difference metric, we observe that Cheng2020 is better than JPEG AI.

\subsection{Effect of channels in chroma branch}
In this experiment, we reduce the number of channels in the chroma branches of SLIC--YUV and SLIC--LAB models from 64 to 32, 16, and 8. This is done in order to understand the effect of colors through the chroma branch. This can be interpreted as the feature space equivalent of color sub-sampling in the image space. For each set of chroma channels, the models are trained. With four bitrate configurations in each model variant, eight RD curves are obtained and shown in Fig.\ \ref{fig:RD_channels}, in which we use JPEG AI as an anchor. We used 100 RGB images of dimensions $400\times400$ from the \emph{Tecknick} dataset \cite{asuni2013testimages}.

Overall, the SLIC--YUV models perform better than SLIC--LAB. The number of channels is directly proportional to the quality. The difference between the variants is higher in PSNR and MS-SSIM. Whereas, in CIEDE2000, both color spaces have a similar performance. JPEG AI outperforms all the variants in terms of MS-SSIM upto a bitrate of 0.5 bpp. After which, the 32 and the 64 channel SLIC--YUV models are better. However, in CIEDE2000, from 16 channel onwards, YUV and LAB variants outperform JPEG AI.

\subsection{Channel impulse responses and color spaces}
The channel impulse responses of a learned image codec provide an insight into the overall features captured by the analysis transform in the form of latent representations. In this experiment, we visualize the impulse responses of the SLIC variants. The impulse response computation is done in the same way as prior work \cite{10222731}. The models with the highest bitrate configuration are chosen. Two images of dimensions $400\times400$ from the \emph{Tecknick} dataset \cite{asuni2013testimages}  are considered here. The impulse responses are first arranged in the decreasing order of their importance by means of their bitrate contribution. The 48 most important channels are considered with the RGB model. In case of YUV and LAB, as there are two branches, the 32 highest channels from the luma branch, and 16 highest from the chroma branch are considered.

The grayscale image \texttt{GRAY\_R03\_0400x0400\_014.png} and its color counterpart \texttt{RGB\_R03\_0400x0400\_014.png} are encoded and reconstructed with all the SLIC variants, and shown in Fig.\  \ref{fig:Gray} and Fig.~\ref{fig:RGB}. The impulse responses for each color space are shown below the reconstructed image. Each patch has dimensions $16\times16$ in the impulse response and represents individual channels in the latent. The luminance impulse responses are shown in the first two rows and the chroma impulse responses are shown in the third row for the LAB and YUV models. The rate and distortion metrics are provided below each image for reference. Comparing the visual quality of reconstructed images, they are very similar, which is also clear from the rate and PSNR values. Irrespective of the model color space, the quality and bitrates are comparable for the example images.

Observing the impulse responses of the grayscale images across all color space variants, it is clear, as one would expect, no color is captured. The main information contained is related to structure, which is represented in the first two rows of SLIC--YUV and SLIC--LAB images in Fig.\ \ref{fig:Gray}. Similar behaviour can be observed for SLIC--RGB, where the top 48 channels resemble structural filters. However, when we observe the impulse responses of the color image counterparts in Fig.\ \ref{fig:RGB}, again we observe a similar behaviour, where the last rows of SLIC--YUV and SLIC--LAB are now populated with colored regions. In case of SLIC--RGB, the impulse responses are a mix of both color and structural components since there is no explicit separation of luminance and chromiance components. For the image considered, since a large part is the blue sky background, this is reflected as the second most important or second highest bitrate contributing channel. 

When we consider the top most channel, and use only this channel to perform the synthesis transform while setting the rest to zero, a low resolution version of the original image is obtained. This means that irrespective of color space, the first channel is most often similar to a low pass filter. The successive channels capture other finer details and colors in the image. Using channel impulse responses, it can be inferred that the number of structural features captured by a learned image codec's non-linear transform is oftentimes higher than that of color features. By separating the luminance and chrominance channels with a color transform, we can have control on the constituents. However, when such a split is not made, deep neural networks learn them implicitly but this results in a lack of optimization and control of the components.

Hence, we can conclude that the features captured by the YUV, LAB and RGB variants of the SLIC model have similarities. With an explicit separation of structure and color, it can be observed that they can be independently optimized and tuned. Whereas, for the RGB model, a granular control is not directly possible with the single branch structure, but training with the loss function in (\ref{eqn:loss}) has an improvement in performance for all variants.
\section{Conclusion}\label{sec:conclusion}

In this paper, we report our findings on the effect of color space in learned image compression. Building on our prior work, we compare the rate-distortion performance of our SLIC model variants with other codecs. It is shown that YUV and LAB models have similar performance. But the RGB model outperforms them at the cost of a higher complexity. The measurements also show that the RGB model has $1.2$ times more number of parameters, and requires $3.2$ times higher kMACs/pixel to that of the YUV and LAB variants. With the channel impulse responses, it is shown that the features captured by color space variants of SLIC have similarities. However, the split model architecture has the benefit of reducing complexity. The experiments can be extended to other color spaces such as HSV, XYZ etc.

\bibliographystyle{IEEEbib}
\small
\bibliography{refs}
\end{document}